\documentclass[aip,amsmath,amsthm,amssymb,citeautoscript,reprint]{revtex4-2}
\usepackage{graphicx, float, xcolor, pgf} 
\usepackage{physics, siunitx}
\usepackage{amsmath}
\usepackage{subfigure}
\usepackage{mathrsfs}
\usepackage{hyperref}
\hypersetup{
    colorlinks=true,
    allcolors=blue,
}

\allowdisplaybreaks
\begin{document}

\title{Excitonic description of singlet fission beyond dimer model : a matrix product state approach}

\author{Supriyo Santra}
\affiliation{School of Chemical Sciences, Indian Association for the Cultivation of Science, Kolkata, India}

\author{Amartya Bose}
\email{amartya.bose@tifr.res.in}
\affiliation{Department of Chemical Science, Tata Institute of Fundamental Research, Mumbai, India}

\author{Debashree Ghosh}
\email{pcdg@iacs.res.in}
\affiliation{School of Chemical Sciences, Indian Association for the Cultivation of Science, Kolkata, India}

\begin{abstract}
    The importance of singlet fission as a fundamental process with a variety of implications in energy harvesting cannot be overstated. The challenge is in characterizing the energy states of these large singlet fission molecular aggregates that participate in the process. Large dimensionality and essential multi-configuration nature of the electronic states of interest combine to make accurate \textit{ab initio} calculations prohibitively difficult. We present a spin-resolved tight-binding excitonic model for singlet fission that can be parameterized based on \textit{ab initio} calculations on monomers and dimers of molecules, and is highly suitable for the study of aggregates using tensor network methods such as the density matrix renormalization group. This tensor network coarse-grained model is demonstrated specifically on the pentacene crystal, where we evaluate the spectra and density of states. We show the natural emergence of bands of states in some cases, and characterize them. Through an analysis of entanglement entropy of the eigenstates, we gain crucial insight into the extent of their multireference character. This method is useful in understanding not just the structure of these extended aggregates, but also being the cornerstone for incorporation of vibronic features and simulation of the singlet fission dynamics.
\end{abstract}

\maketitle

\section{Introduction}
\begin{figure}
    \centering
    \includegraphics[width=0.48\textwidth]{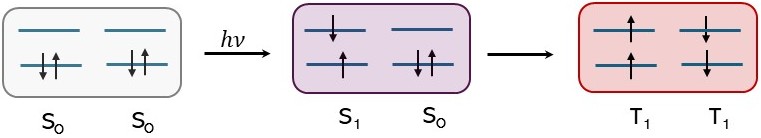}
    \caption{Schematic SF phenomenon}
    \label{fig:sf_scheme}
\end{figure}

Singlet fission (SF) is the phenomenon of creating two separate triplet excitons
on neighboring molecules from a localized exciton created on one
molecule from photoexcitation.\cite{smith2010singlet, smith2013recent} The process happens via an
intermediate state, referred to as the triplet pair state or correlated triplet state
($^1(TT)$ state). This state conserves the spin of the whole system to a singlet
and therefore, the process becomes spin-allowed.\cite{breen2017triplet} 
This first step of SF phenomena is schematically shown in Fig. \ref{fig:sf_scheme}. The process has shown promise
in enhancing the efficiency of Si-based solar cells, and is therefore of
practical importance. \cite{rao2017harnessing, wilson2013singlet}  However,
certain conditions need to be maintained for a molecule to be
SF-active. From an energy standpoint, the monomer's excited bright state energy
should be twice the triplet state energy ($E(S_1)\ge 2\times E(T)$). 
Also, the success of the SF process depends on the presence of significant coupling
between the $^1(TT)$ state and the initial local excited state.
\cite{smith2010singlet, zeng2014low, johnson2013role, yost2014transferable} 

Acenes and carotenoids are two known classes of molecules that show SF. Acenes
have been explored to a great depth from experimental and theoretical
perspectives. The charge transfer (CT) state is shown to be an important intermediate for the formation of the TT-pair state from the initial bright
local excited state (LE).\cite{omar2020elucidating, casanova2018theoretical,
berkelbach2013microscopic2, monahan2015charge} Also, the vibronic aspects of the
processes in acenes have been discussed by several
authors.\cite{andrzejak2018vibronic, kim2021tracking, duan2020intermolecular} 
On the other hand, owing to a complicated excited state manifold in carotenoids, understanding is scarce. Recently, transient absorption
spectra \cite{kundu2021photogeneration} and resonance Raman study
\cite{peng2024investigation} have identified the important role played by the CT
states. In our previous studies, we have shown the structural effect in the SF
process, role of low-lying dark A$_g$ state with CT component and the
vibrational impact in tuning the energetics of the electronic states on a
decapentaene dimer model.\cite{santra2022mechanism, santra2025unraveling}
Several other studies have advanced the understanding of the time scale of the
processes and so on.\cite{barford2023theory,manawadu2023dynamical}

However, the SF process occurs in an aggregate or crystalline form of the molecules, and the majority of the theoretical understanding is based on the dimer model. It is to be noted that,
excitonic state in organic crystals or aggregates may extend to several
chromophores, and therefore an analysis based on dimer models does not
necessarily explain the accurate excitonic photophysics of the
process.\cite{lim2004exciton, casanova2015bright} Using calculations based on
TDDFT and RAS-SCF on tetracene derivatives and rubrene, participation of a minimum of seven monomers in the formation of the excitonic state has been shown.
\cite{casanova2014electronic} This observation was also in line with previous observations from time-resolved spectroscopy on tetracene.\cite{lim2004exciton} 
Similar argument that four monomers are involved in the excitonic state for pentacene was
also made by Head-Gordon and co-workers.\cite{zimmerman2011mechanism} 
By employing QM/MM strategies for pentacene, the important role played by the
environment (and therefore the limitation of the dimer model) in dictating the
SF yield depending on crystal packing was also
highlighted.\cite{suarez2021influence} 

Significant efforts are also made in the development of a Frenkel exciton-like model
Hamiltonians for the singlet fission problem in a diabatic
basis.\cite{berkelbach2013microscopic2,li2020ab} Such approaches have the
advantage of avoiding the scaling problems of accurate electronic structure
methods while, in principle, retaining the accuracy of high-level
electronic structure calculations. These models are most commonly used when
trying to account for the effects of vibronic couplings in dynamics through some
approximate framework. A variety of approaches have been tested for inclusion of
nuclear effects, ranging from Bloch-Redfield master
equations,\cite{ berkelbach2013microscopic1,berkelbach2013microscopic2,berkelbach2014microscopic}
quantum master equation~\cite{nakanoTensorNetworkApproach2021} to tensor
network-based approaches.\cite{pengStudiesNonadiabaticDynamics2023}
On the
other hand, it has also been shown that such model Hamiltonians can be used to
obtain accurate information about the
eigenstates~\cite{zeng2014low}. Such developments
notwithstanding, most studies have been limited to dimers and
occasionally small aggregates. Dynamically, these systems do not allow for the
singlet conformer triplet-pair to diffuse out and for the spin-dephasing to
occur. From a static picture, one does not get the idea of bulk eigenstates that
would be relevant for the process of singlet fission. Consequently, the goal
would be to build on these prior developments and enable the simulation of
the structure and dynamics of large aggregates.

Recently, Li \textit{et al}~\cite{li2020ab} have used a similar
excitonic-model Hamiltonian aimed at capturing the phenomenon of singlet
fission. In this work, they manually extended the Hilbert space, and
correspondingly the Hamiltonian, from the typical dimeric calculations to the
case of a decamer by carefully choosing product basis vectors for the relevant
singlet space. This approach, while novel, is subject to severe constraints in
terms of the number of monomers. Enumerating the basis and the Hamiltonian grows
polynomially in complexity with the number of monomers. Additionally, spin
dephasing processes and vibrational relaxation brings non-singlet states into
the picture as well. These processes are crucially required for the separation
of the entangled triplet pair state to the separate triplets. Therefore, we need
an automated way of producing the Hamiltonian for a large aggregate while
accounting for all the states, including the non-singlet states.

Our current work is an effort in solving this conundrum. Here, we present a
tight-binding-inspired Hamiltonian that is easily representable as a matrix
product operator (MPO) that is parameterized by high-level \textit{ab initio}
calculations. Density matrix renormalization group
(DMRG)~\cite{whiteDensityMatrixFormulation1992,
schollwockDensitymatrixRenormalizationGroup2005,
schollwockDensitymatrixRenormalizationGroup2011, baiardi2020density,
chan2016matrix, ghosh2008orbital, chan2011density} is used to obtain the
variationally optimized eigenstates of the aggregate Hamiltonian. In
Sec.~\ref{sec:method} we discuss the details of this formulation, followed by
the parameterization in Sec.~\ref{sec:electronic-structure}. The
tight-binding-inspired Hamiltonian presented here is dependent on the exact
monomeric states required to properly describe the physics in question. We
discuss the specific implementation for the family of acenes. The numerical
results are presented in Sec.~\ref{sec:results} where we verify the accuracy of
the model and utilize it to study bulk features of these aggregates. We finally
end with some concluding remarks in Sec.~\ref{sec:conclusion}. Singlet fission
as a process is intrinsically dependent on time-dynamics. In this discussion we
also provide a discussion on how the MPO Hamiltonian is a critical step in
incorporating this spin dephasing process and vibronics.

\section{Method}\label{sec:method}

Direct electronic structure calculations on singlet fission aggregates become
unfeasible for large aggregates. We approach the problem through a combination
developing a Frenkel exciton-like model Hamiltonian parameterized by accurate
electronic structure theory and subsequent use of DMRG to obtain eigenstates and
properties for aggregates.

We develop and demonstrate the method for the family of acenes, and pentacene in
particular. For singlet fission in aggregates of acenes, the monomeric states of
interest are well-known to be the ground state ($\ket{G}$), the singlet excited
($\ket{S}$), the triplet state and the cationic and the anionic
states~\cite{berkelbach2013microscopic2,
zeng2014low}. However, to
be able to account for the spin dephasing process, one needs to spin resolve this
monomeric basis. In addition to the ground and excited states, which being spin
0 configurations have no spin variants, we need to include the
two spin-resolved cationic ($\ket{C_\uparrow}$ and $\ket{C_\downarrow}$) and
anionic ($\ket{A_\uparrow}$ and $\ket{A_\downarrow}$) states, and the three
triplet states ($\ket{T_{+1}}$, $\ket{T_0}$, and $\ket{T_{-1}}$). The basis of direct product
states constructed out of these states will span the entire Hilbert space
for the aggregate. Each of the states with multiple spin types would be denoted
with two indices, the first for the spin and the second for the monomer number. So,
$\ket{T_{sj}}$ would imply the triplet state with $s$ spin and on the $j$th
site. The ground and the singlet excited states, which do not have different
spin variants, would only have a single index corresponding to the monomer
number.

The Hamiltonian can be expressed in terms of these states. To motivate the
Hamiltonian, let us first understand the structure of the terms:
\begin{align}
    \hat{H} &= \hat{H}_\text{site} + \hat{H}_\text{corr} + \hat{H}_\text{hopping} + \hat{H}_{S\rightarrow C} + \hat{H}_{S\rightarrow A}\nonumber\\
    &+ \hat{H}_{S\rightarrow TT} + \hat{H}_{CA\rightarrow TT} + \hat{H}_{AC\rightarrow TT}\label{eq:Hamiltonian}
\end{align}
where $\hat{H}_\text{site}$ are the terms in the Hamiltonian which give the
one-body energies of the various states on each of the monomers,
$\hat{H}_\text{corr}$ gives the two-body correlation energies like the
triplet-pair correlation and the correlation of the charge transfer states. The
hopping of different excitations across the chain without losing their character
is described by $\hat{H}_\text{hopping}$.

Furthermore, for notational simplicity, it is most convenient to introduce
notations for some two-body states in terms of the one-body basis. Since we are
interested in the singlet sector of the spectrum, which arises out of
interactions between singlet direct-product states. Additionally, the
interactions will all be two-body terms. So, we introduce short-hand notation
for the two-body singlet spin conformers of charge-transfer states and the
triplet-pair states.
\begin{align}
    \ket{^1\left(CA\right)_{j,k}} &= \frac{1}{\sqrt{2}}\left(\ket{C_{\uparrow j}A_{\downarrow k}} - \ket{C_{\downarrow j}A_{\uparrow k}}\right)\label{eq:CA}\\
    \ket{^1\left(AC\right)_{j,k}} &= \frac{1}{\sqrt{2}}\left(\ket{A_{\uparrow j}C_{\downarrow k}} - \ket{A_{\downarrow j}C_{\uparrow k}}\right)\label{eq:AC}\\
    \ket{^1\left(TT\right)_{j,k}} &= \frac{1}{\sqrt{3}}\left(\ket{T_{+1j}T_{-1k}} + \ket{T_{-1j}T_{+1k}} - \ket{T_{0j}T_{0k}}\right)
\end{align}

Using these states, we can provide the detailed operator forms of each of
the terms in the Hamiltonian.
\begin{widetext}
\begin{align}
    \hat{H}_\text{site} &= \sum_j \left(\epsilon^{(G)}\dyad{G_j} + \epsilon_j^{(S)} \dyad{S_j} + \epsilon_j^{(T)}\sum_s\dyad{T_{sj}} + \epsilon_j^{(C)}\sum_s\dyad{C_{sj}} + \epsilon_j^{(A)}\sum_s\dyad{A_{sj}}\right)\label{eq:site-en}\\
    \hat{H}_\text{corr} &= \sum_{j<k} \sum_{s, s'}\left(\epsilon^{(CA)}_\abs{j-k} \dyad{C_{sj}A_{s'k}} + \epsilon^{(AC)}_\abs{j-k} \dyad{A_{s'j}C_{sk}}\right)\nonumber\\
    &+\sum_{j<k} \epsilon^{(TT)}_\abs{j-k} \sum_{s,s'}\dyad{T_{sj}T_{s'k}}\nonumber\\
    &+\sum_{j<k} \epsilon^{(GG)}_\abs{j-k} \dyad{G_jG_k}\nonumber\\
    &+\sum_{j<k} \epsilon^{(SG)}_\abs{j-k} \dyad{S_j G_k} + \sum_{j<k} \epsilon^{(GS)}_\abs{j-k} \dyad{G_j S_k}\label{eq:corr}\\
    \hat{H}_\text{hopping} &= \sum_j h^{(S)}\left(\dyad{S_jG_{j+1}}{G_jS_{j+1}} + \text{h.c.}\right) + \sum_j h^{(T)}\sum_s\left(\dyad{T_{sj}G_{j+1}}{G_jT_{s,j+1}} + \text{h.c.}\right)\label{eq:hopping} \\
    \hat{H}_{S\rightarrow C} &= \sum_j h^{(SG\rightarrow CA)} \dyad{S_jG_{j+1}}{^1\left(CA\right)_{j, j+1}} + h.c.\label{eq:sg_ca}\\
    \hat{H}_{S\rightarrow A} &= \sum_j h^{(SG\rightarrow AC)} \dyad{S_jG_{j+1}}{^1\left(AC\right)_{j, j+1}} + h.c.\\
    \hat{H}_{S\rightarrow TT} &= \sum_j h^{(SG\rightarrow TT)} \dyad{S_jG_{j+1}}{^1\left(TT\right)_{j,j+1}} + \sum_j h^{(GS\rightarrow TT)} \dyad{G_jS_{j+1}}{^1\left(TT\right)_{j,j+1}} + h.c.\\
    \hat{H}_{CA\rightarrow TT} &= \sum_j h^{(CA\rightarrow TT)} \dyad{^1\left(CA\right)_{j,j+1}}{^1\left(TT\right)_{j,j+1}} + h.c.\\
    \hat{H}_{AC\rightarrow TT} &= \sum_j h^{(AC\rightarrow TT)} \dyad{^1\left(AC\right)_{j,j+1}}{^1\left(TT\right)_{j,j+1}} + h.c.\label{eq:ac_tt}
\end{align}
\end{widetext}
It can be seen that, that in all the hopping terms, Eqs~\ref{eq:hopping} to~\ref{eq:ac_tt},
the coupling coefficients are nearest neighbor. The tensor network formalism
does not demand this, and for increasing computational cost can handle
long-distance couplings as well. In fact, we are already incorporating
long-distance the correlation term, Eq.~\ref{eq:corr}. However, it shall be
demonstrated in Sec.~\ref{sec:electronic-structure}, the coupling coefficients
turn out to be nearest neighbor along all the directions.

We also point out at this stage that depending on the exact
definitions of the monomeric states, it is possible that the ground state diabat
$\ket{G_1G_2\ldots G_N}$ may be coupled with other states~\cite{zeng2014low}.
This would imply that the true ground state of the aggregate is not identical to
this diabat basis. Our definition of the Hamiltonian has currently ignored such
possibilities for simplicity. However, it is possible to include such effects
through terms like $\dyad{S_jG_{j+1}}{G_jG_{j+1}} + h.c.$

The size of the Hilbert space is $d^N$ where $d$ is the number of monomeric
states considered and $N$ is the number of monomers. In this case, $d=9$. This
exponential scaling makes direct computations of large aggregates unfeasible. We
represent the states of the aggregates as a matrix product state.

We utilize DMRG for obtaining the eigenstates of the Hamiltonian in a
variational manner. The number of eigenstates grows exponentially with
the size of the Hilbert space, and the accuracy and efficiency of variational
methods suffer for highly excited eigenstates. Moreover, even if one could
extract all the eigenstates, it is difficult to filter out the ones that are
relevant for any process out of the near continuum of eigenstates that would be
generated. The singlet fission process starts with a single monomer that gets
Frank-Condon excited to $\ket{S}$, while all the other monomers remain in the
ground state. We are interested in understanding the spectrum of eigenstates
that are connected with this singly-excited state,
and thereby restrict the variational search for the states of interest.

Imagine that the initial excitation is on monomer number $j$,
$\ket{\psi(0)} =
\prod_{k=1}^{j-1}\ket{G_k}\otimes\ket{S_j}\otimes\prod_{l=j+1}^{N}\ket{G_l}$.
The wave function satisfies the Schr\"odinger equation,
\begin{align}
    i\hbar\pdv{\ket{\psi(t)}}{t} &= H\ket{\psi(t)},
\end{align}
with the Hamiltonian in Eq.~\ref{eq:Hamiltonian}. As the initial state evolves
in time, it would span a subspace of the full Hilbert space. We are interested
in the eigenstates that span the same subspace. The Hamiltonian connects
$\ket{\psi(0)}$ to all other states with a single singlet excitation on arbitrary
monomers. Additionally, these states are also connected to charge-separated
states where there is a pair of cationic and anionic monomers, or states with
two triplets. Notice that this subspace of the aggregate Hilbert space is
characterized by being charge neutral. This gives us the first condition
satisfied by the eigenstates that we are interested in. Charge neutrality has to be preserved.

A second condition also emerges from the time-dependent analysis. Notice that
just like the eigenstates with local charges have pairs of oppositely charged
local states, the triplets that emerge with time evolution also
occur in pairs. As long as the initial condition is a single locally excited
$\ket{S}$ state, then it can lead to only states that either have one $\ket{S}$,
or a pair of $\ket{C}$ and $\ket{A}$ states, or a pair of $\ket{T}$ states.
These states, in the absence of spin-dephasing, is guaranteed to be singlet in
nature, and consequently will have $S_z=0$. Ideally, a stronger statement is to
say that the total $S^2$ should be a conserved quantity, but we avoid this for
simplicity as it is not additive over the tensor network sites. Therefore,
tensor networks would be defined in terms of states with charge $Q$, and $S_z$
as quantum numbers. It should be highlighted here that by using just these two
conditions, we will get a superset of eigenstates, including some that are not
singlet states at all. One can naively imagine doing a post-processing and
filtering out all the non-singlet aggregate eigenstates.

\begin{figure}
    \centering
    \includegraphics[width=0.24\textwidth]{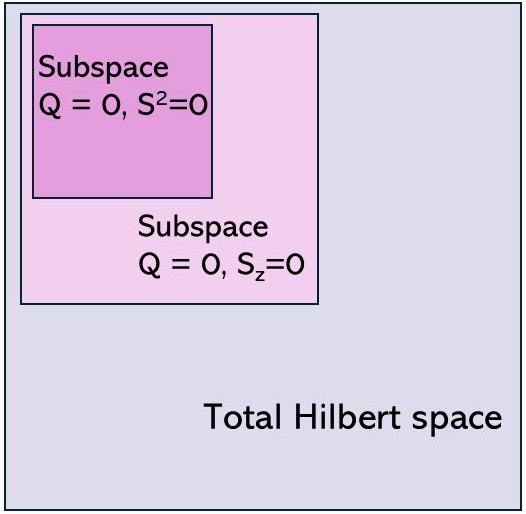}
    \caption{Schematic representation of an important subspace of the total Hilbert space characterized by the charge quantum number, $Q=0$ and $S^2 = 0$. This is an important space for the SF phenomenon.}
    \label{fig:H_subspace}
\end{figure}

We can now utilize DMRG~\cite{whiteDensityMatrixFormulation1992,
schollwockDensitymatrixRenormalizationGroup2005} that is restricted to the
states satisfying these two conditions to obtain the eigenstate spectrum. All
tensor network calculations were implemented using the ITensor
library~\cite{fishmanITensorSoftwareLibrary2020,
fishmanITensorSoftwareLibrary2022}. We start the variational optimization in the
$(Q=0, S_z=0)$ subspace of the Hilbert space. This gives us the first excited
state. Subsequent excited states are generated by ensuring that they are
orthogonal to the previously calculated states while remaining in the same
subspace. This orthonormality is ensured by adding to the original Hamiltonian,
projectors to each of the states that the new state has to be orthogonal to,
with some energy penalty. It is not trivial to estimate the number of such
eigenstates. However, we know that there cannot be more eigenstates than the
dimension of the $(Q=0, S_z=0)$ space. Once we exceed that number, the
orthogonality condition would necessarily be violated. We use this as a flag to
denote the completion of the calculation of eigenstates. While the cardinality of the full Hilbert space grows exponentially in the number of monomers ($9^N$ in this case owing to the 9-dimensional monomeric basis), the cardinality of the $Q=0, S_z=0$ subspace grows merely polynomially as $\mathcal{O}(N^2)$. This is schematically shown in Fig.~\ref{fig:H_subspace}.

The problem with doing a filtration in terms of the value of $S^2$ after
generating the superset of states constrained by only $Q=0$ and $S_z=0$ is that
one has to do a significantly greater number of DMRG runs. The cost of these
runs increases with the number of sites. Additionally, in the absence of more
specific constraints, there is a greater spin contamination in the resultant
states. One can avoid this post-processing step by noticing that in the super
set of states generated, the excess states would have $S^2>0$ with $S_z=0$. This
is guaranteed by the quantum numbers. However, if we can produce a complete
basis of states that have $S^2>0$ and $S_z=0$, and ensure the orthonormality of the
DMRG eigenfunctions to this set, then one would be directly left with the
singlet states. We look at all possible states where all but two monomers are in
the ground state, and these two monomers are in a two-body state such that they
satisfy $S_z=0$. Given the space that defines the monomeric or diabatic basis,
we notice that the cation and anion states, which are doublets, can come
together to give states of $S^2=1, S_z=0$ or a triplet state. Similarly, two
triplets can come together to give states with $S_z=0$ but $S^2 = 1$ or 2. These
bimolecular triplet and quintet states need to be avoided. (There are no other
spin states for these effectively bimolecular wavefunctions because of the
nature of the monomeric basis for acenes.) Once these states are generated using
the Clebsch-Gordon coefficients, we add them to a set of states that the
targeting eigenstates need to be orthonormal to. This procedure eliminates the
need for filtration using $S^2$ and also provides absolute spin-purity of the
resultant eigenstates.

Once all the eigenstates are obtained in the site-local basis in the form of
MPS, it is trivial to generate the absorption spectrum. For acenes, the
monomeric $\ket{S}$ states are bright. Thus, it is possible to describe the
transition dipole operator,
\begin{align}
    \vec{\hat{\mu}} &= \sum_j \vec{\mu}_j \left(\dyad{S_j}{G_j} + \dyad{G_j}{S_j}\right),
\end{align}
as a matrix product operator. Using this operator, we generate the stick absorption spectrum of the aggregates with the intensity being proportional to $\abs{\mel{\psi_j}{\vec{\hat{\mu}}}{\psi_g}}^2$.

Further, every state is associated with its character, which determines the
singlet excitation component, the triplet excitation, and the charge transfer
components in the state. The various populations are calculated as,
\begin{align}
    S_\text{pop} &= \sum_j \braket{\psi}{S_j}\braket{S_j}{\psi}\\
    T_\text{pop} &= \sum_j\sum_s \braket{\psi}{T_{js}}\braket{T_{js}}{\psi}\\
    CT_\text{pop} &= \sum_j\sum_s \braket{\psi}{C_{js}}\braket{C_{js}}{\psi}.
\end{align}
Notice that the charge transfer population can either be calculated as the
cationic population or the anionic population because of the $Q=0$ condition
that every state satisfies. Following this, the $S$-character is defined as
$\frac{S_\text{pop}}{S_\text{pop} + T_\text{pop} + CT_\text{pop}} \times 100
\%$. The other characters are defined in analogous manners. In every plot, we
assign a color to the character by mapping the $T$-character, $S$-character, and
the $CT$-character onto a red-green-blue (RGB) color space.

\section{Parameters for constructing the Hamiltonian}\label{sec:electronic-structure}

\begin{figure}[!htb]
    \centering
    \subfigure[]{\includegraphics[width=0.24\textwidth]{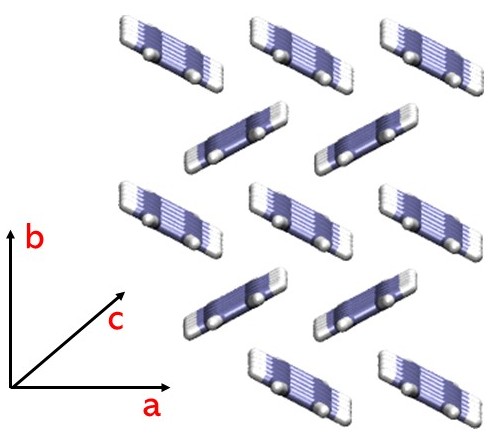}}
    \subfigure[]{\includegraphics[width=0.35\textwidth]{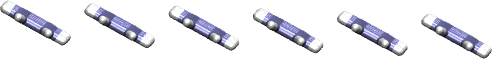}} 
    \subfigure[]{\includegraphics[width=0.35\textwidth]{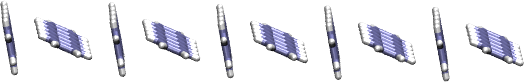}}
\caption{(a)The crystal structure from the Ref.\citenum{schiefer2007determination}. The crystallographic axes are shown (b) Aggregate structure along a-axis, (c) Aggregate along the Herringbone or zigzag direction}
    \label{fig:crys_str}
\end{figure}

The parameters for site energies and the interactions between the monomers are
obtained from the accurate \textit{ab initio} calculation on the monomer and dimer structures
taken out from the crystal thin film structure as reported in Ref.
\citenum{schiefer2007determination} (shown in Fig. \ref {fig:crys_str}a). We focus our discussion on the two prominent directions, $\vec{a}$ (Fig.~\ref{fig:crys_str}b) and the herringbone direction (Fig.~\ref{fig:crys_str}c).

The S$_1$ state for the dimer is
driven by HOMO $\rightarrow$ LUMO transition. For parameterizing the Hamiltonian in Eq.~\ref{eq:Hamiltonian}, in addition to the one-body terms, simulations of dimers are also required. Therefore, for the dimeric calculations of
the two-body parameters, we have taken an active
space of (4e,4o) comprising HOMOs and LUMOs on each monomer. This suffices
to describe all the required states necessary for SF in the case of acenes.

Calculations were performed on the dimer with varying intermonomer distances, yielding the low-lying adiabatic states. These are reported for both geometries in the SI (Fig.~S1). Of the 5 excited states calculated we see that the S$_1$ is a TT-pair state while S$_2$ and S$_3$ are dominated LE character. S$_4$ and S$_5$ adibatic states are CT states at small separation.  These adiabats, however, are not the direct inputs for
Eq.~\ref{eq:Hamiltonian}, which is written in the diabatic basis. The first step,
therefore, is to diabatize the CASSCF molecular orbitals.

\begin{figure}[!htb]
    \centering
    \subfigure[Along $\vec{a}$ direction]{\includegraphics{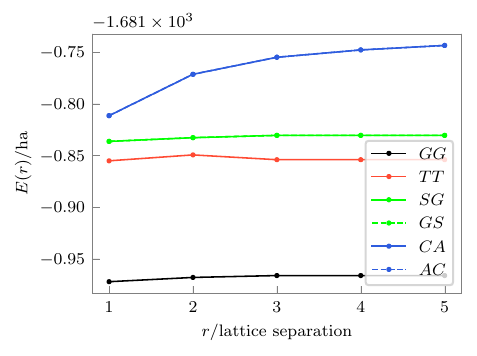}}
    \subfigure[Along herringbone direction]{\includegraphics{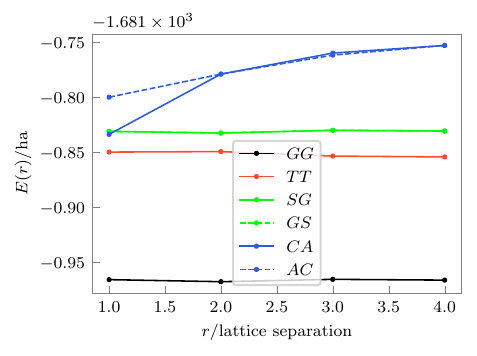}} 
    \caption{Energies of the dimeric diabatic states as a
function of separation distance (a) along $\vec{a}$ and (b) along herringbone orientation}
    \label{fig:dimer_ener_comp}
\end{figure}  

After localization of the canonical CASSCF orbitals using the Pipek-Mezey scheme~\cite{pipek1989fast}, we obtain the diabatic basis. In Fig.~\ref{fig:dimer_ener_comp}, we show the distance dependence of the energies of various diabats. Notice that the correlations in the energies of most of these diabats die off within a separation of 2--3 units. However, the distance dependence of the charge transfer diabats, $\ket{CA}$ and $\ket{AC}$, is somewhat long-lived. This is to be expected because the asymptotic dependence for these diabats should be the same as Coulomb's interaction, decreasing as $1/r$. For most of the cases, the left-right symmetry is preserved --- for example the energy of the $\ket{SG}$ state is the same as the energy of the $\ket{GS}$ state. The only exception is the large difference in energy of the $\ket{CA}$ and $\ket{AC}$ states at the nearest neighbor herringbone conformation. This owes its origin to multipolar effects generated from asymmetric charge distributions due to the particular conformation in question. From these curves, and using some of the monomeric calculations, it is relatively simple to obtain all the site-based energies, Eq.~\ref{eq:site-en}, and the correlation terms, Eq.~\ref{eq:corr}.

First, we discuss the site-based energy terms. While the
diabatic energies can of course be obtained from the monomeric calculations,
they can also be derived from the dimeric calculation. This is the route we take to
ensure greater consistency. To illustrate this, consider the energy of the $GG$
diabat as a function of energy, $E_{GG}(r) = \mel{GG}{\hat{H}}{GG}$. This is the
black line in Fig.~\ref{fig:dimer_ener_comp}a. At long distances, the interaction
between the two monomers goes to zero. Consequently, $\lim_{r\to\infty} E_{GG}(r)
= 2\epsilon^{(G)}$. The same argument is used to obtain the value of
$\epsilon^{(T)}$. Now the singlet excitation energy, $\epsilon^{(S)}$ can be
obtained as $\lim_{r\to\infty} E_{SG}(r)-\epsilon^{(G)}$. The cationic energy
$\epsilon^{(C)}$ is obtained from the monomeric calculations. Finally, to obtain
the energy of the anionic state, we fit the charge-transfer diabatic energy to
$E^{(CT)}_{\infty} + E^{(CT)}_\text{corr}/r$. The anionic value is then obtained as
$\epsilon^{(A)} = E^{(CT)}_\infty - \epsilon^{(C)}$.

The correlation terms of Eq.~\ref{eq:corr} are now obtained by
subtracting the long-distance value of the energy from the corresponding
diabatic energy. For instance, $\epsilon^{(GG)}_r = E_{GG}(r) - 2\epsilon^{(G)}$
or $\epsilon^{(TT)}_r = E_{TT}(r) - 2\epsilon^{(T)}$. The $SG$ correlation is
obtained as $\epsilon^{(SG)}_r = E_{SG}(r) - \epsilon^{(S)} - \epsilon^{(G)}$.
The last charge-transfer correlation terms are obtained by subtracting
$E^{(CT)}_\infty$.

For calculation of the coupling and hopping terms (i.e., $h^{(i)}$s in Eqn \ref{eq:hopping} to \ref{eq:ac_tt}), we also need the localized orbital description of the dimer states. For the dimer, if $\ket{\psi_i}$ and $\ket{\psi_f}$ represent the two diabatic states, then the coupling between them is given as $\bra{\psi_i}\hat H\ket{\psi_f}$. The expression of the coupling terms between several diabatic states is taken from the references \citenum{santra2022mechanism,berkelbach2013microscopic2}. 
For example, the singlet hopping term $h^{(S)}$ is expressed as $\bra{SG}\hat H\ket{GS}$, where one singlet exciton has hopped from the monomer on the left to the right. These coupling and hopping terms also decay very rapidly with lattice distance. 

It should be emphasized that both the diagonal correlations and the off-diagonal coupling are extremely sensitive to the orientation of the monomers. Till now, our discussion has focused on the direction $\vec{a}$ of the lattice, which is the simplest. To demonstrate the dependence on orientation, we consider the herringbone direction (shown in Fig. \ref{fig:crys_str}c). All the methods of analysis remain the same, but the actual values show enormous differences. On cursory inspection of Fig.~\ref{fig:dimer_ener_comp}, it can be seen that the $\ket{AC}$ and $\ket{CA}$ diabats show the most prominent difference. Unlike the parallel dimer (Fig.~\ref{fig:dimer_ener_comp}a), for the herringbone dimer, there is significant splitting between $\ket{AC}$ and $\ket{CA}$ diabats 
(Fig.~\ref{fig:dimer_ener_comp}b). This arises because of the asymmetry of the monomers in herringbone orientation. This brings the $\ket{AC}$ diabat below the local excited states, which has implications in the SF mechanism.  This splitting between the $\ket{AC}$ and $\ket{CA}$ diabats is absent in the parallel dimer because of the symmetric packing of the monomers.     
Since each odd entry for the herringbone direction will have this inherent asymmetry in the dimer structure, slight splitting of the diabat states can be observed. Detailed methods for obtaining the parameters and the magnitudes of them are given in the SI.  

Along directions $\vec{b}$ and $\vec{c}$, the monomers are arranged parallely while having a much larger intermonomer distance than $\vec{a}$. Therefore, we do not expect qualitatively different results from those of $\vec{a}$ oligomers. We do not expect the CT states to separate in these directions and, therefore, expect these directions to be relatively less important for the SF process. The same has been implicitly used in many earlier theoretical descriptions of SF phenomena, where the dimers along the zigzag direction have been used as the only ones of importance.\cite{zeng2014low,berkelbach2013microscopic2,zimmerman2011mechanism,zimmerman2010singlet,yost2014transferable}

\section{Results and discussion}\label{sec:results}

As we have obtained all the parameters to build the Hamiltonian, we systematically employ our tensor network-based method for obtaining molecular properties. 
\subsection{Excited states of dimer}\label{sub_sec:excited}
We have computed the low-lying adiabatic singlet excited states of the
dimer along the direction $\vec{a}$, and compared our results with \textit{ab initio}
calculation (6SA-(4e,4o)-CASSCF/6-31G(d)). This provides a test-bed for our
method.

\begin{table}[h]
\centering
\caption{Comparison of the excited states from our method and (4e,4o)-CASSCF calculation along $\vec{a}$ direction.}
\begin{tabular}{c|cc|cc}
\hline
State & \multicolumn{2}{c|}{Our method} & \multicolumn{2}{c}{SA-6-(4e,4o)-CASSCF} \\
\cline{2-5}
       & VEE (eV) & Nature & VEE (eV) & Nature \\
\hline
S$_1$      & 3.16 (0.00)     & TT       & 3.55 (0.00)      & TT      \\
S$_2$      & 3.56 (0.60)      & LE      & 3.80 (0.79)     & LE      \\
S$_3$      & 3.80  (0.00)    & LE      & 3.96 (0.00)      & LE      \\
S$_4$      & 4.38   (0.00)   & CT      & 4.73  (0.00)     & CT      \\
S$_5$      & 4.38    (0.00)  & CT      & 4.73   (0.00)   & CT      \\
\hline
\end{tabular}
\label{tab:comp_a}
\end{table}

Table \ref{tab:comp_a} compares the vertical excitation energies (VEE) and electronic character of low-lying excited states obtained from our method and the SA-6-(4e,4o)-CASSCF/6-31G(d) calculation. A consistent agreement is observed in the nature of the states across the two approaches, with all excited states retaining the same character. The S$_1$ state is the TT-pair state. Both S$_2$ and S$_3$ are LE states, resulting from the combination of 
$\ket{SG}$ and $\ket{GS}$ diabats. Similarly, the S$_4$ and S$_5$ are CT states comprising of $\ket{AC}$ and $\ket{CA}$ diabats. We obtain the nature of the states by analyzing the quantities $S_{pop},T_{pop}$ and
$CT_{pop}$ defined in section \ref{sec:method}. 
The VEEs are also in good agreement with each other. These discrepancies arise primarily due to the incomplete set of diabats considered in forming our Hamiltonian. Upon verification of the adiabatic states obtained from the \textit{ab initio} method, we see several more configurations within the
(4e,4o) active space also contributes in constituting the adiabatic states. In particular, for the S$_1$ state, contributions were observed from multiexcitonic diabats. 
These are not included in our method. However, despite this obvious simplification, the restricted set of diabats captures the essential physics, including the nature of the states and the degeneracy of the S$_4$ and S$_5$ states. 

The absorption spectrum along $\vec{a}$ obtained using our tensor network method is shown in SI and the VEEs with oscillator strengths are shown in Table~\ref{tab:comp_a}. 
In the spectrum, we additionally use color to denote the nature of the state --- a local excitation $S$-like state is colored green, a CT state is colored blue, and a triplet-like state is colored red. Fractional contributions are colored by combinations of the primary colors. This spectrum shows one clear dominant peak corresponding to the $S_2$ state at around $\SI{3600}{\meV}$. The \textit{ab initio} calculations have the same characteristics with a single dominant peak. A much weaker feature appears at higher energy (\SI{4400}{\meV}), assigned to a pure CT state (blue), which carries negligible oscillator strength. The absence of any other transitions implies no mixing between LE and CT states, consistent with the well-separated adiabatic states observed in Table~\ref{tab:comp_a}. 
Such a spectral profile reflects a relatively weak electronic coupling in the parallel geometry, where orbital overlap is minimal.

In the context of SF, $\vec{a}$ is less significant due to the minimal coupling. It is well documented that the SF happens in acenes via the CT state.~\cite{berkelbach2013microscopic2}  On the other hand, it is well-known that the herringbone orientation of the pentacene dimer plays critical role in the process. We have already seen in section~\ref{sec:electronic-structure} that the diabatic parameters change quite significantly in comparison to direction-$\vec{a}$. This difference in parameters naturally gets reflected in the eigenenergies and the dimeric adiabatic wave functions.

Table~\ref{tab:comp_z} presents a comparison of the low-lying excited-state manifold of the pentacene dimer in the herringbone orientation, with the absorption spectrum in SI. As in the case of the parallel dimer, there is consistent agreement in the ordering and electronic character of the states between our method and the \textit{ab initio} calculations. The vertical excitation energies (VEEs) are captured with qualitative accuracy.
 
\begin{table}
\centering
\caption{Comparison of the excited states from our method and (4e,4o)-CASSCF calculation along herringbone orientation}
\begin{tabular}{c|cc|cc}
\hline
State & \multicolumn{2}{c|}{Our method} & \multicolumn{2}{c}{SA-6-(4e,4o)-CASSCF} \\
\cline{2-5}
       & VEE (eV) & Nature & VEE (eV) & Nature \\
\hline
S$_1$      & 3.14 (0.00)      & TT      & 3.53 (0.00)      & TT      \\
S$_2$      & 3.48 (0.18)      & LE+CT      & 3.81 (0.38)     & LE+CT      \\
S$_3$      & 3.64 (0.14)     & LE      & 3.86 (0.16)      & LE      \\
S$_4$      & 3.80  (0.28)    & LE+CT      & 4.07 (0.15)      & LE+CT      \\
S$_5$      & 4.55  (0.00)    & CT      & 4.90 (0.01)      & CT      \\
\hline
\end{tabular}
\label{tab:comp_z}
\end{table}

As shown in Table~\ref{tab:comp_z}, the excited states of the pentacene dimer in the herringbone orientation exhibit significant mixing between LE and CT character across nearly the entire low-lying manifold. Therefore all the states have moderate optical brightness. This is in contrast to the parallel dimer configuration, where the states retain more distinct and
well-separated character. This enhanced LE-CT mixing in the herringbone geometry can be understood as a result of the near-degeneracy of the $\ket{SG}$ / $\ket{GS}$ diabat and the $\ket{CA}$ diabat shown in Fig.~\ref{fig:dimer_ener_comp}~(b). Mixed characters of eigenstates result from a combination of coupling between the diabatic states and their energies being close. While in the case of the parallel geometry, the energies of the locally excited diabats and the CT states never come close, that is not the case for the herringbone geometry leading to greater mixing. Furthermore, this mixing is also enhanced by the asymmetry and orbital overlap inherent in the
molecular packing. In particular, such mixing facilitates coupling between bright LE states and multiexcitonic configurations through intermediate CT states, thereby potentially enhancing the efficiency of the SF process. All the coupling and other parameters are given in SI.

\begin{figure*}[!htb]
    \centering
    \subfigure[Spectrum along  $\vec{a}$]{\includegraphics[]{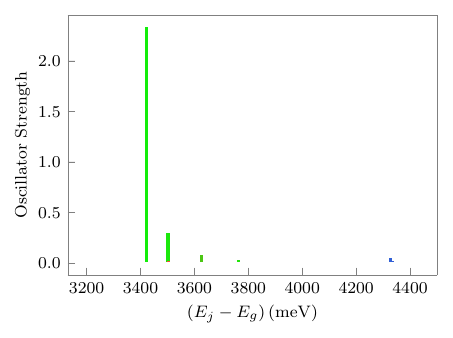}}
    \subfigure[State density along  $\vec{a}$]{\includegraphics[]{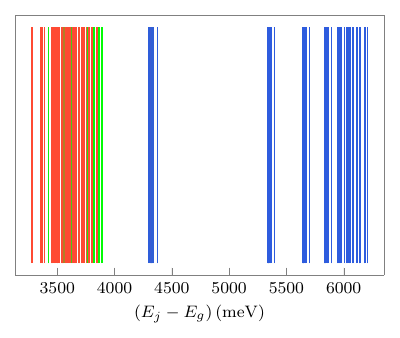}}
    \subfigure[Spectrum along herringbone orientation]{\includegraphics[]{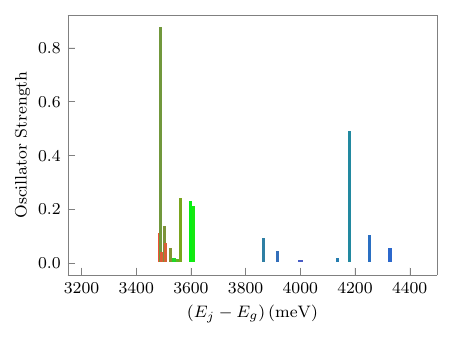}}
    \subfigure[State density along herringbone orientation]{\includegraphics[]{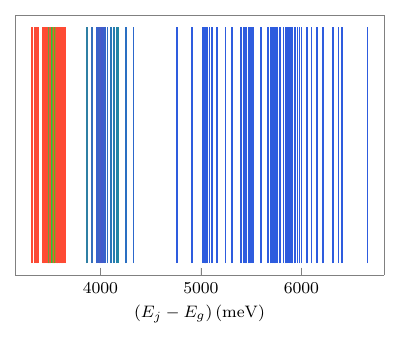}}
\caption{Spectra of bulk (a) and state densities (b) for parallel aggregate. Spectra of bulk (c) and state densities (d) for herringbone aggregate.}
    \label{fig:bulk_spec_state}
\end{figure*}  

\subsection{Aggregate absorption spectra and state densities}

Having established the validity of the tensor network approach using the dimer,
we move towards modeling the larger aggregates. We construct an extended
Hamiltonian by incorporating additional sites into the effective model, using
parameters derived from the dimer level \textit{ab initio} calculations. We have
computed the spectra and state densities for hexamer and heptamer which shows
that the spectra converges within the system size of 6-monomers. Therefore, for
the final results we have computed the spectra and state densities of a decamer
as the effective aggregate spectra. Fig.~\ref{fig:bulk_spec_state} shows the
decamer spectra and state densities along $\vec{a}$ and zigzag directions. Both
the spectra and state densities show remarkable differences in both these
directions.

From the spectrum of the bulk along $\vec{a}$ (Fig.~\ref{fig:bulk_spec_state}a),
one bright peak at 3441 meV is observed, which is almost a pure LE state
(96 \%).  As we increase the system size for the parallel dimers,  we observe a
slight red shift of this intense peak. Another smaller peak emerges around 3600
meV from trimer onwards, which also has predominant LE character. Therefore, the
spectra has not changed much in the bulk compared to the dimer spectra along
$\vec{a}$.  Furthermore, since the monomers are all oriented in the same way, the
dipole moment of each monomer just add up for the bright state reflecting the
increased brightness of the peak.   

We show the state density of the aggregate in the
Fig.~\ref{fig:bulk_spec_state}b, that shows that the states remain almost
entirely pure (no mixing of different nature of the states) as evident from the
pure blue, red and green colors for the states. Furthermore, we see that there
are natural gaps arising between the CT states and LE states and they are well
separated. This is further proof that the lack of strong coupling between
monomers along $\vec{a}$ leads to its relative lack of importance in SF
phenomena. The separation within the CT states is noteworthy. On analysis
of the state character we observe that the lowest energy CT-band (from 4300 -
4380 meV) consisted of CT states on immediate monomers. The next band, at almost
1000 meV apart, are the cluster of states having CT character on next nearest
neighbors. Similarly, as the monomers having ionic characters are further
separated, the energy keeps increasing and forms several bands after that.  This
is not surprising in the context of Coulomb interactions. Nearer the two
oppositely charged bodies, greater is the stabilization. Hence, we see that
states with the CT nature on immediate neighbors form the most red shifted (or
stable) band among all the CT-bands. 

Next, we turn our focus on the herringbone orientation of the monomers and
inspect the spectral features. As expected from an extrapolation of the dimeric
spectrum, we observe many more bright states and often in relatively dense
clusters (Fig.~\ref{fig:bulk_spec_state}~(c)). Two most intense peaks are seen
around 3490 meV having a mixed nature (23\% S + 67\% T + 10\% CT). These intense
transitions are also surrounded by several moderate peaks dominated by LE
character. However, significant amount of mixing of the CT character is found
in the bright states obtained around 3500 meV (indicated by dark red peaks, combination of red, blue and green). This
mixed nature involving the LE and CT states would increase the efficiency of SF
phenomenon in herringbone structure. The CT states are quite significantly
spread out and the formation of bands that appeared along $\vec{a}$ is
not observed for the herringbone aggregates
(Fig.~\ref{fig:bulk_spec_state}~(d)). Additionally, in the broad band of mostly
CT-like states from $\SIrange{3860}{4330}{\meV}$, we observe a varying degree of
blue colors arising from the mixing of LE (green) character. These CT states reside on nearest neighbor monomers. The next few CT states would have charges residing on next nearest neighbor and so on.
However, unlike the parallel aggregate the
CT states are spread out in zigzag orientation and do not form clear bands. 
This might be attributed to the left-right asymmetry of the CT states upon swapping (i.e.,
the asymmetry of charge distribution and energy between $\ket{CA}$ and
$\ket{AC}$).

\begin{figure}
    \centering
    \subfigure[$\vec{a}$ Axis]{\includegraphics{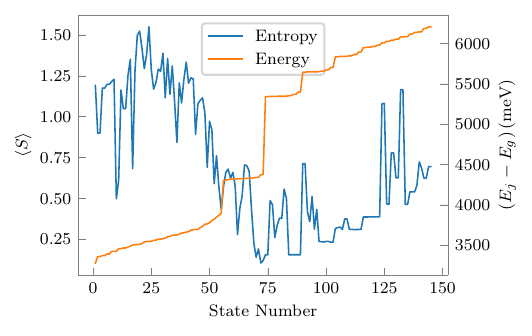}}
    
    \subfigure[Herringbone Axis]{\includegraphics{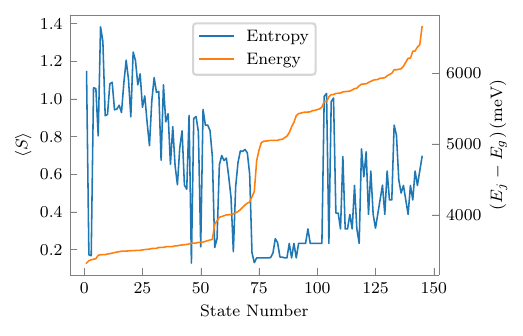}}
    \caption{Average entanglement entropy (calculated across the MPS on the excitonic basis) for the different states along both the directions. The energy of the states are denoted in orange.}
    \label{fig:entanglement-entropy}
\end{figure}

Next we inquire into the entanglement entropy of the eigenstates. Each of the
eigenstates are represented as an MPS. The entanglement entropy corresponding to
a bond measures the entanglement between the two subsystems that bond divides
the whole system into. We show the the entanglement entropy averaged across all
the bonds for every eigenstate for both the directions in
Fig.~\ref{fig:entanglement-entropy}. Details of this calculation are given in SI. Though the exact entanglement entropy of
the two geometries are different, the trends look similar. The entanglement entropy of the low-lying states up to 60--70 are high, followed by a region
between roughly the 70th and the 100th states where the entanglement entropy is
exceptionally low. After that the average entanglement entropy starts to
increase slightly but still remains low. So, it seems that purely CT states have
low entanglement entropy. The lowest values of the CT state cannot be zero
because every CT state has to have a singlet conformer. The spin combinations
that yields this for a dimer are already seen in Eqs.~\ref{eq:CA}
and~\ref{eq:AC}, and these are Bell-like entangled states.

This feature can also be envisaged in the context of the
predominantly single-reference nature of CT states, where in an excitonic basis,
CT states are represented as fragment-localized product states composed of
cationic and anionic site states. In contrast, LE states in molecular aggregates
inherently exhibit multireference character, as they require linear combinations
of multiple local excitation configurations to capture excitonic
delocalization.~\cite{pitesa2024excitonic, li2017ab, li2020ab} This \textit{multireference} is due to the near degeneracy in the excitonic basis. This manifests in high entanglement entropy on the occupation number of the excitonic basis. On the other hand, the CT states lead to charge separation that leads to the environment of the charged species encountering different fields and therefore, localization along the excitonic basis, i.e., the charges on the charged monomers are located on specific monomers only. This leads to low entanglement entropy.

The energy dependence of the states shows a step-wise increase (Fig.~\ref{fig:entanglement-entropy}). 
The first jump around state
number 55 or so. The second and generally higher jump in energy occurs around state number 75 and this
corresponds to the energy spacing between the predominantly LE or mixed LE-CT states and the purely CT states. This has implications in the subsequent dynamics of excitons during singlet fission phenomena.
Though the current work is not about dynamics, this entire analysis
sheds light on expectations that one should have when time evolution is taken
into account. If we are starting with an excitation into the bright manifold,
probably the states that we would be most concerned about is the band near
$\SI{3600}{\meV}$ or the CT-like band upto \SI{4400}{\meV}. Both these bands are
energetically isolated from the extremely high lying CT states. Moreover, the
entanglement entropy of those states is also extremely low. This isolation of
the bright bands from the high lying CT states imply that irrespective of the
non-adiabatic couplings between these adiabats, the dynamics would likely be
restricted to the manifold below the separation between the CT bands. These
ideas would be fully developed in the future once the vibrations are also
incorporated through appropriate means.

\section{Conclusion}\label{sec:conclusion}

In this paper, we have demonstrated that a parametrized form of the aggregate
Hamiltonian can be solved using DMRG to obtain important information about the
the low-lying electronic states and absorption spectra of singlet fission
aggregates. Pentacene has been explored as a particular example system.

For the tight-binding model, one has to decide upon a diabatic or site-local
basis to represent the problem. To ensure both continuity with previous
literature and simplicity, we have chosen the ground state $\ket{G}$, the
singlet excited state $\ket{S}$, the three triplet states $\ket{T_s}$ and the
two cationic $\ket{C_s}$ and anionic $\ket{A_s}$ states as our basis. This is an
extension of previous models~\cite{berkelbach2013microscopic1,zeng2014low,li2020ab}
 used to study singlet fission in the
pentacene dimers. Those previous studies have created truncated dimeric diabatic
basis of interest from the five type of states. In our work, we augment the
traditional basis to incorporate the value of $S_z$. This is important in order
to describe the entanglement of the triplet-pair state for instance, and would,
in the future, be crucial for describing the dynamical loss of this entanglement
through vibronic couplings.

The tight-binding Hamiltonian is expressed as a matrix product operator quantized with the total charge and the value of $S_z$ as quantum numbers. DMRG was used to calculate the eigenstates in the $Q=0, S_z=0$ sub space. These are additionally projected to yield only the singlet states. In a similar vein, the transition dipole moment operator is also written in the form of a vector of MPOs. The matrix elements connecting the ground state to a particular eigenstate is used to obtain the amplitude of that state.

With this machinery in place, we test our model Hamiltonian parameterized at CASSCF level of theory with the adiabats obtained directly from dimeric \textit{ab initio} calculations. Following verification of the qualitative features, we demonstrated the efficiency of the method in trying to understand the structure of larger aggregates. We predict and explore the spectra of aggregates along the $\vec{a}$ and herringbone directions in the bulk limit.

The particular diabatic basis used to describe an aggregate is highly dependent
on the system and phenomena at hand. While for this specific problem, we have chosen the basis accordingly, one could choose a different optimal set of basis sets and Hamiltonian to understand a different phenomena.
Our CASSCF calculations on the dimeric system at various
distances reveal significant contributions from monomeric states with
multi-excitonic character or di-ionic character. Including the spin-resolved
versions of these states would bring higher accuracy to the tensor network
results for the aggregates but at a much greater computational cost. Future work
would focus on incorporating more elaborate monomeric Hilbert spaces. Furthermore, the ultimate goal would be to use these effective Hamiltonians and knowledge from the resulting spectrum into a complete dynamical description of the phenomena. Work in this direction is also underway.

\section{Acknowledgement}
S.S. thanks DST-Inspire for fellowship and IACS for computational facilities.
D.G. thanks ANRF-CRG grant (CRG/2023/001806) for funding. A.B. thanks TIFR for
providing access to computational facilities.

\bibliography{ref}
\end{document}


\author{Supriyo Santra$^1$, Amartya Bose$^{\star2}$, and Debashree Ghosh$^{\star1}$}
\date{%
$^\star$Email: amartya.bose@tifr.res.in, pcdg@iacs.res.in 
\\$^1$School of Chemical Sciences, Indian Association for the Cultivation of Science, Kolkata 700032, India\\
$^2$Department of Chemical Sciences, Tata Institute of Fundamental Research, Mumbai 400005,
India
}
\maketitle
\section{Adiabatic excited states of dimers }
We have computed the low-lying excited states of the dimer. For the claulctaion we have taken the (4e,4o) active space and used the state-averaged complete active space self consistent (SA-CASSCF) field method using a 6-31G(d) basis set. Quantum chemistry software Molpro was used for these claculation \cite{werner2015molpro}. As mentioned in the manuscript here S$_1$ state is the $^1(TT)$-pair state having contributions also from the multi-excitonic configurations. S$_2$ and S$_3$ states are local excited states having the combination of $\ket{SG}$ and $\ket{GS}$ diabats. At short distances, the S$_4$ and S$_5$ states are dominated mainly by CT states dominated by $\ket{AC}$ and $\ket{CA}$ diabats.

\begin{figure}[!htb]
    \centering
    \subfigure[along $\vec{a}$]{\includegraphics[width=0.45\textwidth]{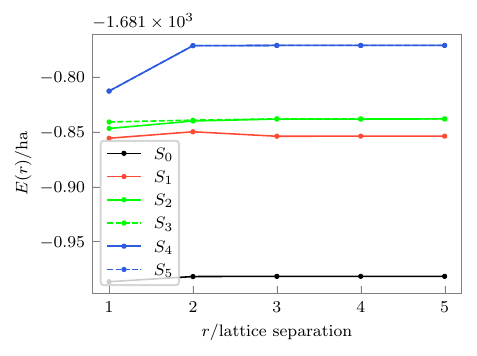}} 
    \subfigure[along herringbone direction]{\includegraphics[width=0.45\textwidth]{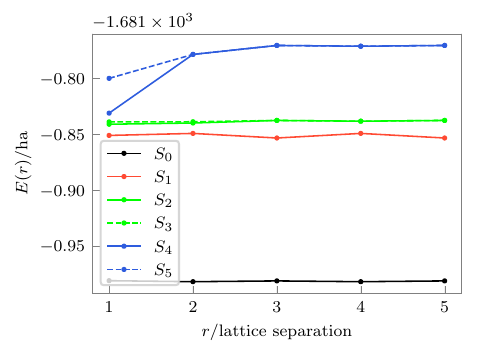}}
 \caption{ Low-lying adiabatic excited states of the dimer as a function of lattice distance}\label{fig:dira-dist}
\end{figure}

As shown in Fig.~\ref{fig:dira-dist}, the orientation of monomers influences the energy ordering of the states. In the herringbone direction, we see that the one of the CT states, has come close to the LE state at short distance. It has significant implication in 
the success of SF process. 

However, the parameters cannot be directly extracted from these adiabatic states. This is not only because our Hamiltonian described in the manuscript is a diabatic one and therefore requires parameters from the states described in diabatic basis, but also the character of the adibatic states changes along the lattice separation. As an illustration it can be said that the adiabatic states (S$_4$ and S$_5$) become significantly multiexcitonic with large separation in contrast to CT nature at short distance.  

\section{Parameterizing the Hamiltonian}\label{sec:parameters}
 In this section, our method and strategies to obtain the parameters are elaborated. 
 
 To construct the Hamiltonian given in the manuscript, we need two kinds of parameters,
\begin{enumerate}
    \item One body parameters (site energies, $\epsilon^{(G)}$, $\epsilon^{(T)}$, $\epsilon^{(C)}$, $\epsilon^{(A)}$) 
    \item Two body parameters (correlation terms, hopping terms, and coupling terms)
\end{enumerate} 
Here one body and two body refers to monomer and dimer. We discuss how we obtain the parameters from 
accurate \textit{ab initio} calculation on monomers and dimers. But for better consistency, we have obtained 
the one body terms also from dimer calculation as explained below. 

\begin{figure}[!htb]
    \centering
    \subfigure[(HOMO)$_m$]{\includegraphics[width=0.22\textwidth]{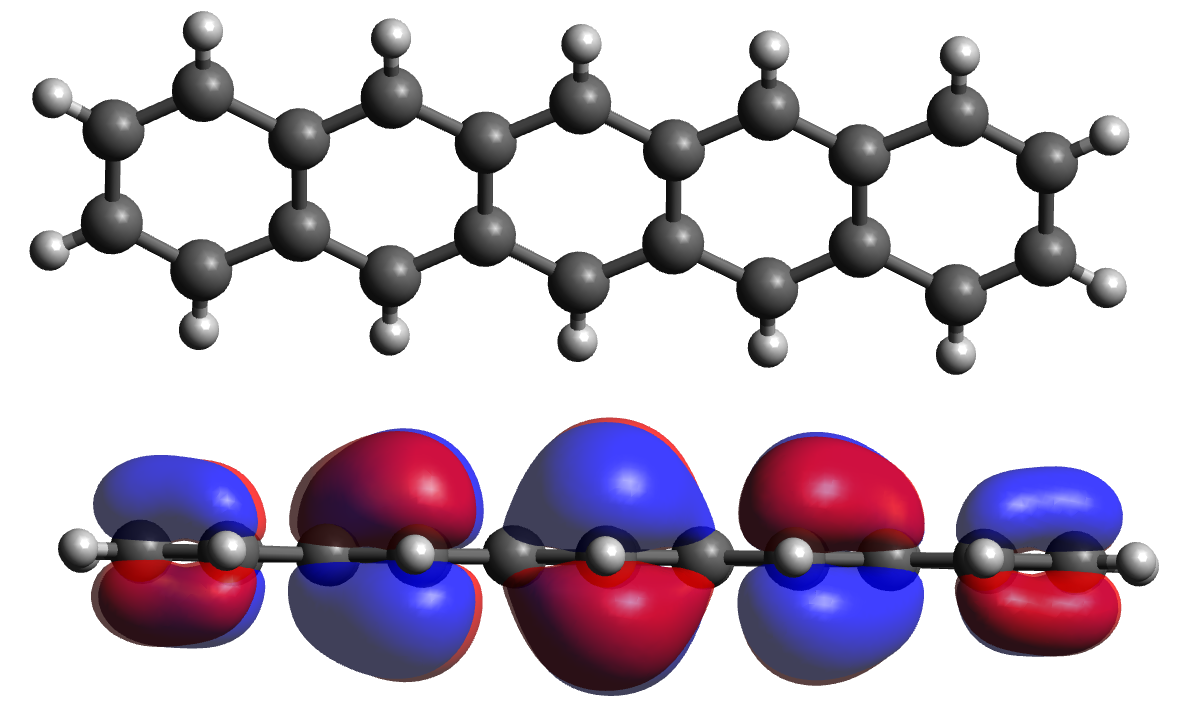}}
    \subfigure[(HOMO)$_n$]{\includegraphics[width=0.22\textwidth]{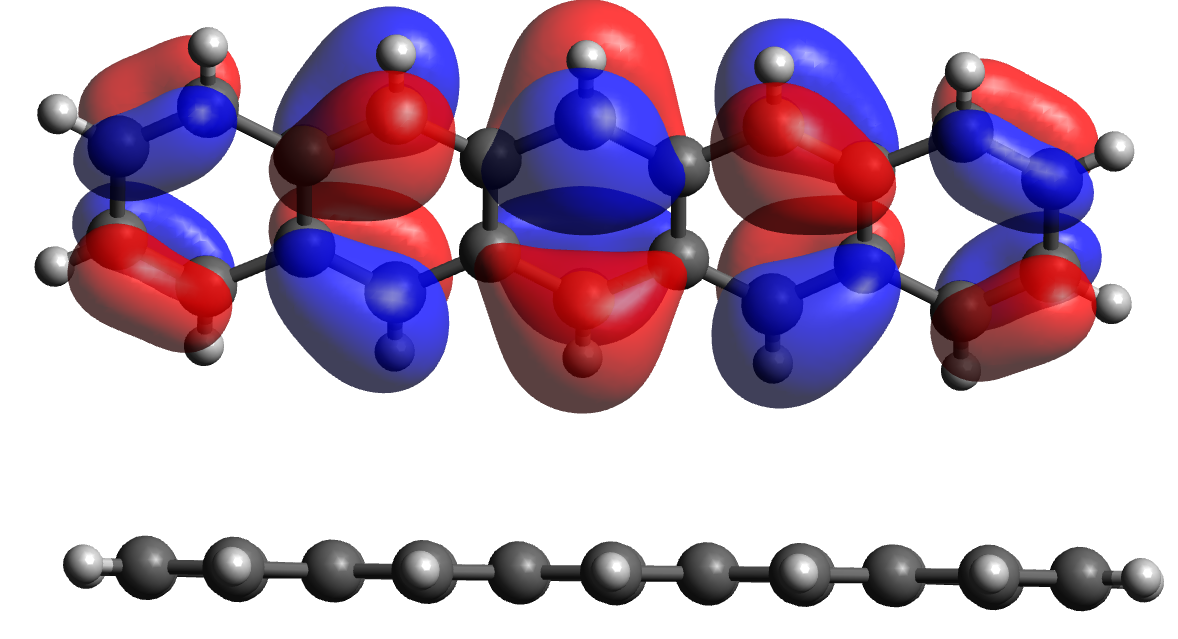}}
    \subfigure[(LUMO)$_m$]{\includegraphics[width=0.22\textwidth]{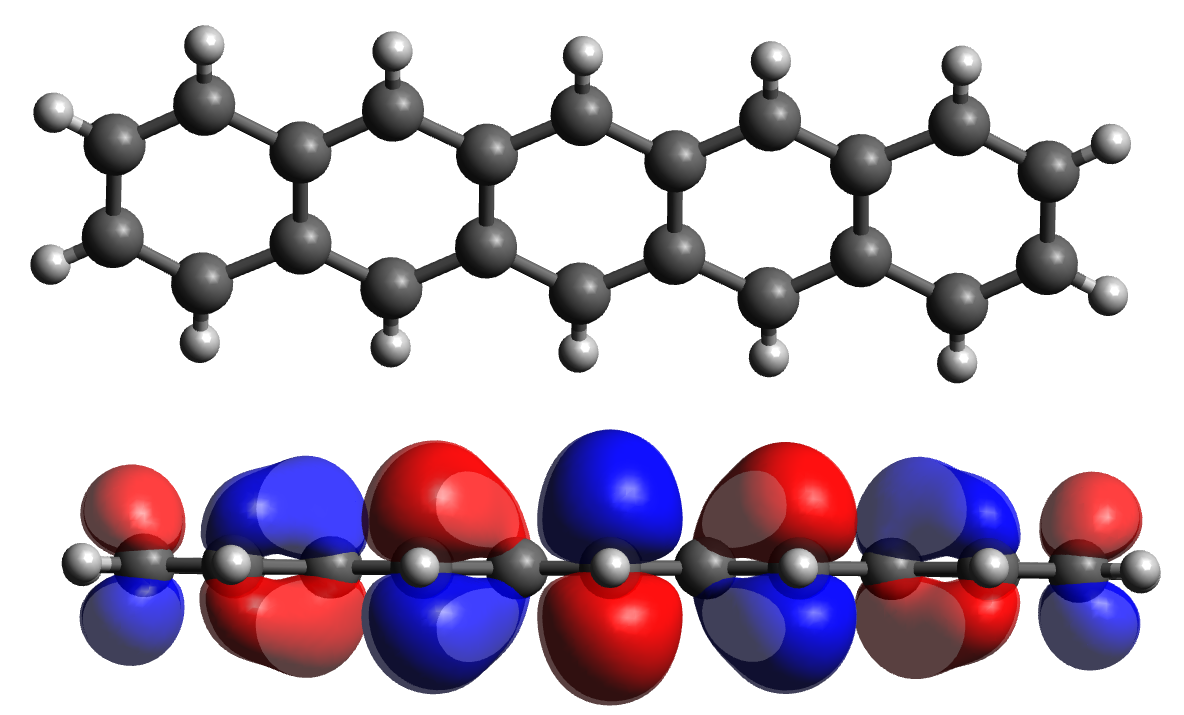}}
    \subfigure[(LUMO)$_n$]{\includegraphics[width=0.22\textwidth]{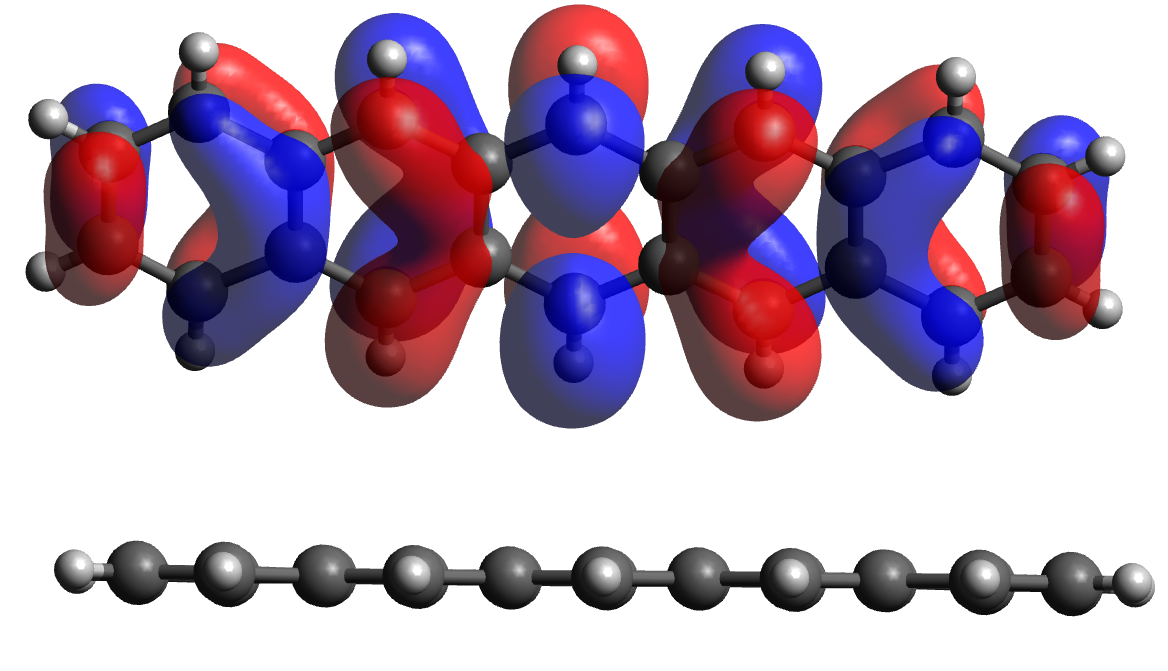}}
 \caption{Localised (diabatic) orbitals, where m and n are the indices of the two monomers}\label{fig:loc_orb}
\end{figure}

Before we discuss how we obtained the parameters from diabatic states, it is appropriate to mention how we construct the diabatic basis. The diabatic basis is  a localised basis on the monomers. For that we take the (4e,4o) CASSCF canonical orbitals and localize them by using Pipek-Mezey scheme. The localised (diabatic) spatial orbitals are shown in the Fig.~\ref{fig:loc_orb}. We have only shown the orbitals along the herringbone direction. Notice that the localization scheme effectively gives us the HOMO and the LUMO of the two monomers. Thus we have motivated a basis whose spatial part is made up of the HOMO and the LUMO of the monomer in question.

\begin{figure}[!htb]
    \centering
    \includegraphics[width=0.80\textwidth]{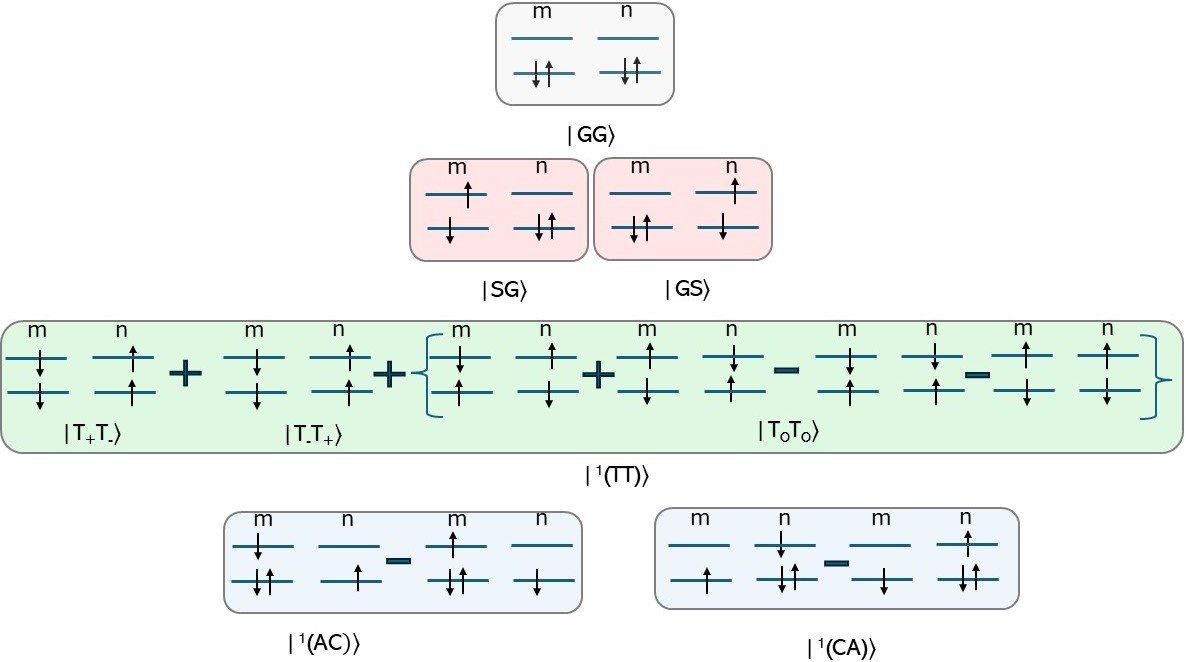}
    \caption{Diabatic states for dimer}
    \label{fig:diab_dimer}
\end{figure}
The Hamiltonian in the tensor network approach is defined on a spin-resolved monomeric basis. We start defining this basis in terms of the CAS determinants. Considering the 2 spatial orbitals discussed above, every monomeric CAS Slater-determinant is written in occupation number notation as $\ket{\ket{o_Ho_L}}$ where $o_H$ and $o_L$ are the populations of the HOMO and LUMO orbitals respectively. Now, we can begin to map our monomeric basis on to this notation as follows:
\begin{itemize}
    \item $\ket{\ket{20}}$ representing the ground state $\ket{G}$, showing that the HOMO of the monomer is doubly occupied.
    \item $\frac{\ket{\ket{\uparrow\downarrow}} - \ket{\ket{\downarrow\uparrow}}}{\sqrt{2}}$ as the singlet-excited state $\ket{S}$.
    \item $\ket{\ket{\uparrow 0}}$ and $\ket{\ket{\downarrow 0}}$ are the two spin-resolved cationic states $\ket{C_\uparrow}$ and $\ket{C_\downarrow}$ respectively.
    \item $\ket{\ket{2\uparrow}}$ and $\ket{\ket{2\downarrow}}$ are the two spin-resolved anionic states $\ket{A_\uparrow}$ and $\ket{A_\downarrow}$ respectively.
    \item There are three triplet states. $\ket{\ket{\uparrow\uparrow}}$ and $\ket{\ket{\downarrow\downarrow}}$ are $\ket{T_{+1}}$ and $\ket{T_{-1}}$ respectively. $\ket{T_0}$ is equivalent to the $\frac{\ket{\ket{\uparrow\downarrow}} + \ket{\ket{\downarrow\uparrow}}}{\sqrt{2}}$.
\end{itemize}

The CAS results are decomposed in terms of various charge neutral, singlet dimeric diabatic states obtained by combining two monomers in this basis --- $\ket{GG}$, $\ket{SG}$, $\ket{GS}$, $\ket{^1\left(TT\right)}$, $\ket{^1\left(CA\right)}$, and $\ket{^1\left(AC\right)}$. While the kets $\ket{SG}$, $\ket{GS}$, and the ground state $\ket{GG}$ are simply the direct products of the monomeric basis defined above, the two charge transfer states, and the triplet pair state are more elaborate entangled linear combinations that are reported in the body of the paper. The dimeric diabatic state electron configurations are schematically shown in Fig.~\ref{fig:diab_dimer}.

\subsection{One-body parameters}
Consider a situtation when the inter monomer separation is large enough that the
dimer state can be expressed as  the product state of the monomers. So, a dimer
diabatic  $\ket{GG}$ state energy($\lim{_{r\to\infty}}E_{GG}(r)$) at that
separation is essentially 2$\times \epsilon_G$(the monomer $\ket{G}$ state
energy). Therefore, we obtain  $\epsilon^{(G)}$ as
$\lim{_{r\to\infty}}E_{GG}(r)/2$. Similarly at the large inter-monomer
separation, we have $\lim _{r\to\infty}E_{GS} = \epsilon_G + \epsilon_S$. Since,
we have already obtained the $\epsilon^{(G)}$, we can estimate $\epsilon^{(S)}$.

A similar though more involved line of argument can be extended for the dimer $\ket{^1\left(TT\right)}$ state. Notice that 
\begin{align}
    \ket{^1\left(TT\right)} = \frac{1}{\sqrt{3}}\left(\ket{T_{m,+1}T_{n,-1}} + \ket{T_{m,-1}T_{n,+1}} - \ket{T_{m,0}T_{n,0}}\right)
\end{align}
The energy of this dimeric diabatic state can be written as:

\begin{align}
E_{TT}(\abs{m-n}) &= \frac{1}{3} \Big( 
    \mel{T_{m,+1}T_{n,-1}}{\hat{H}}{T_{m,+1}T_{n,-1}} 
    + \mel{T_{m,-1}T_{n,+1}}{\hat{H}}{T_{m,-1}T_{n,+1}} 
    + \mel{T_{m,0}T_{n,0}}{\hat{H}}{T_{m,0}T_{n,0}} \nonumber \\
    &\quad + \mel{T_{m,+1}T_{n,-1}}{\hat{H}}{T_{m,-1}T_{n,+1}} 
    - \mel{T_{m,+1}T_{n,-1}}{\hat{H}}{T_{m,0}T_{n,0}} \nonumber \\
    &\quad + \mel{T_{m,-1}T_{n,+1}}{\hat{H}}{T_{m,+1}T_{n,-1}} 
    - \mel{T_{m,-1}T_{n,+1}}{\hat{H}}{T_{m,0}T_{n,0}} \nonumber \\
    &\quad - \mel{T_{m,0}T_{n,0}}{\hat{H}}{T_{m,-1}T_{n,+1}}
    - \mel{T_{m,0}T_{n,0}}{\hat{H}}{T_{m,+1}T_{n,-1}}\Big)
\end{align}

In the infinite separation limit, two things happen. Firstly, all the off-diagonal terms in Eq.... become zero because the monomeric triplet states are eigenstates of the molecular Hamiltonian. Second, the diagonal terms can be written in terms of the one-body energies:
\begin{align}
\mel{T_{m,+1}T_{n,-1}}{\hat{H}}{T_{m,+1}T_{n,-1}} &= \mel{T_{m,+1}}{\hat{H}_{\text{1body}}}{T_{m,+1}} + \mel{T_{n,-1}}{\hat{H}_{\text{1body}}}{T_{n,-1}}\\
\lim_{\abs{m-n}\to\infty}E_{TT}(\abs{m-n}) &= \frac{2}{3}\left(\mel{T_{+1}}{H_\text{1body}}{T_{+1}}+\mel{T_{-1}}{H_\text{1body}}{T_{-1}}+\mel{T_{0}}{H_\text{1body}}{T_{0}}\right)
\end{align}
Now, utilizing the fact that in absence of external magnetic fields, the three triplet states would be degenerate, we can say that the energy of the triplet is $\epsilon^{(T)} = \lim_{r\to\infty}E_{TT}(r)/2$.

However, as the $\ket{AC}$ and $\ket{CA}$ states show stronger dependency to the distance in between the monomers, we fit them to $1/r$ and take the monomer $\epsilon^{(C)}$ from a monomer calculation. Therefore, the $\epsilon^{(A)}$ energy can be obtained by subtracting $\epsilon^{(C)}$ from $E_{CT}(\infty)$. 

\subsection{Two-body parameters}
For obtaining the correlation terms, we have computed the energies of the relevant diabatic states of the dimers as shown in Fig.~\ref{fig:diab_dimer}. The diabatic energies are plotted as function of inter lattice separation, shown in Fig. 4 in the main manuscript. Now for all the diabat states except for $\ket{AC}$ and $\ket{CA}$, we observe that the energy does not change beyond the nearest neighbor. Therefore the correlation vanishes at that separation (i.e, it behaves as a product state at $\infty$ separation). We get the $\epsilon^{(ij)}_{\abs{j-k}}$ where ${ij} = {GG},{SG},{GS},{TT}$. For $\ket{AC}$ and $\ket{CA}$ diabatic states we have considered the $1/r$ dependency of the correlation term, as this is long range correlation and does not die at next nearest neighbor.   

Coupling terms are the off-diagonal elements of the electronic Hamiltonian ($\hat{H_{el}}$) of the dimer in the diabatic basis. If $\psi_i$ and $\psi_f$ are the two diabatic states of the dimer, then the coupling between the two are given as, $\langle\psi_i|\hat{H}_{el}|\psi_f\rangle$.

The electronic Hamiltonian, in second quantization is given as,
\begin{equation}
    \widehat{H}_{el} = \sum_{ij,\sigma} h_{ij}c_{i,\sigma}^\dagger c_{j,\sigma} + \frac{1}{2} \sum_{ijkl,\sigma\sigma^\prime} V_{ijkl}c_{i,\sigma}^\dagger c_{j,\sigma^\prime} c_{l,\sigma^\prime} c_{k,\sigma}^\dagger
    \end{equation}
where the sum runs over all the orbitals and $i,j,k$, and $l$ are the indices of the orbitals on both the monomers. 
The one and two electron integrals are given as, 

 \begin{eqnarray}
        h_{ij} = \int d^3r\phi^* (r) \Big[-\frac{1}{2}\nabla^2 + V_{e-n}(r)\Big]\phi_j(r)\equiv(i|\hat{h}|j)\nonumber\\
        V_{ijkl} = \int d^3r_1\int d^3r_2 \phi_i^*(r_1)\phi_j(r_1)r_{12}^{-1} \phi_k^*(r_2)\phi_l(r_2)\equiv(ij|kl)\nonumber
\end{eqnarray}
respectively. 
The one and two-electron integrals are again obtained using the Molpro quantum chemistry package. The expression of the coupling terms are taken from the references \cite{berkelbach2013microscopic2,santra2022mechanism}.

\subsection{Values of the parameters}

We present the one-body and two-body couplings parameters obtained as described above in Tables~\ref{tab:one-body-params} and~\ref{tab:two-body-params} for both the directions. These parameters are in Hartree. 
\begin{multicols}{2}
    \begin{centering}
        \begin{tabular}{c|c|c}
            Parameter & Direction $\vec{a}$ & Herringbone \\\hline
            $\epsilon^{(G)}$ & -840.9831 & -840.9831\\
            $\epsilon^{(T)}$ &-840.9270 & -840.9270 \\
            $\epsilon^{(S)}$ & -840.8474& -840.8474\\
            $\epsilon^{(C)}$ & -840.7882& -840.7882\\
            $\epsilon^{(A)}$ & -840.9387 & -840.9361\\
        \end{tabular}
        \captionof{table}{One-Body Parameters}
        \label{tab:one-body-params}
    \end{centering}
    \columnbreak
    \begin{centering}
        \begin{tabular}{c|c|c}
            Parameter & Direction $\vec{a}$ & Herringbone \\\hline
            $h^{(S)}$ &-0.0042 & 0.0012 \\
            $h^{(T)}$ &-0.0021 & 0.0006\\
            $h^{(SG\rightarrow CA)}$ &-0.0024 & -0.0044 \\
            $h^{(SG\rightarrow AC)}$ & -0.0024& -0.0039\\
            $h^{(SG\rightarrow TT)}$ & -3.70$\times 10^{-7}$& -9.05$\times 10^{-6}$ \\
            $h^{(GS\rightarrow TT)}$ & 3.70$\times 10^{-7}$& 7.63$\times 10^{-6}$\\
            $h^{(CA\rightarrow TT)}$ & -0.0037 & 0.0031\\
            $h^{(AC\rightarrow TT)}$ & 0.0028& -0.0024\\
        \end{tabular}
        \captionof{table}{Two-Body Coupling Parameters}
        \label{tab:two-body-params}
    \end{centering}
\end{multicols}

Notice that though the one-body parameters were fit separately for the $\vec{a}$
direction and the herringbone direction, the values are extremely similar. The
coupling parameters can, in principle, be long distance as well. However, the
values become zero after the nearest-neighbor term. One also sees this data
recovering the well-known facet that SF in acenes is never direct, and always
mediated by the charge transfer states. All the direct couplings between the
$\ket{SG}$-like states and the triplet-pair state are zero.

The long-range correlations that exist in the diabatic energies can directly be
seen from the diabatic energy plots provided in the main paper.

\section{Details of the Density Matrix Renormalization Group Calculations}
After obtaining the parameters for the Hamiltonian in the diabatic basis as
described in section~\ref{sec:parameters} and the body of the paper, a matrix
product operator is formed. This Hamiltonian is optimized to obtain the ground
state using DMRG. Further more, excited states are obtained sequentially by
variationally optimizing them while maintaining orthonormality with the
previously calculated states. Every step of DMRG is repeated 5 times and the
lowest energy solution is kept in order to prevent the algorithm from getting
stuck in a local minimum. The orthonormality to a precomputed set of states,
$\left\{\psi_j\right\}$, is ensured by adding a penalty term to the Hamiltonian
as follows:
\begin{align}
    \hat{\tilde{H}} &= \hat{H} + \sum_j w_j \dyad{\psi_j},\label{eq:excited}
\end{align}
where $\ket{\psi_0}$ is the ground state, $\ket{\psi_1}$ is the first excited
state and so on. Now, the major question that needs to be answered is how does
one know when to stop. One option is to put an upper bound on the energy of the
eigenstates that we want to calculate. In this work, we take a different
approach. Notice that the space that we are doing the DMRG simulations in is
characterized by $Q=0$ and $S_z=0$. This is a finite space, say with a
dimensionality of $D$. There can only be $D$ orthonormal vectors that span this
space. Therefore, when we try to optimize for the $(D+1)$th eigenstate, the
penalty term in Eq.~\ref{eq:excited} will not be able to ensure the satisfaction
for the orthonormality condition. The overlap of the wave function obtained from
DMRG will have substantial overlap with one or more of the previously calculated
eigenstates. This signals that the space has already been completely described
and can be used as a stopping rule for the search for eigenstates.

Every DMRG run is done with 50 sweeps, with cutoffs increasing from $10^{-20}$
to $10^{-30}$ in 5 steps and remaining constant, and the maximum bond dimension
increasing from 200 to 1500 in 8 steps and remaining fixed. A noise of
$10^{-15}$ was used.

\section{Average Entanglement Entropy of MPS Eigenstates}
\begin{figure}
    \centering
    \subfigure[Direction $\vec{a}$]{\includegraphics{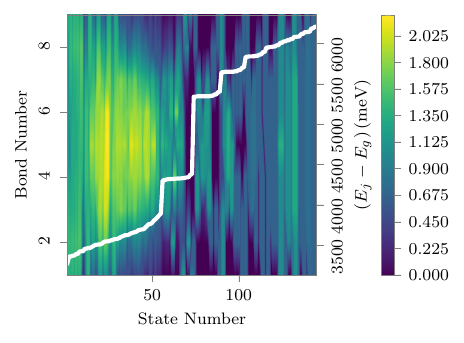}}
    ~\subfigure[Herringbone direction]{\includegraphics{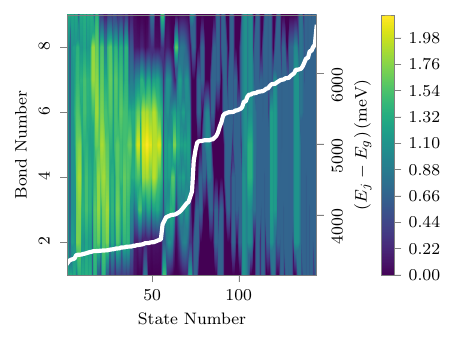}}
    \caption{Bond resolved entanglement entropy for every state along different directions. White bold line shows the energy of that particular state.}
    \label{fig:entanglement-entropy}
\end{figure}
In the paper we have reported the average entanglement entropy for each excited
eigenstate. In this section, we discuss the definition of this quantity.
Entanglement entropy is a bipartite entropy measuring the entanglement between
two subsystems. For an eigenstate defined as an MPS, every bond index that
connects the site-based tensors provide a location for partitioning the MPS into
two subsystems. Suppose we partition the system into $b$ sites in the left
subsystem and $N-b$ sites in the right subsystem. For this partitioning one can
calculate the entanglement entropy in the following manner:
\begin{enumerate}
    \item Set the orthogonality center of the MPS on the $b$th site.
    \item Perform the singular value decomposition of the $b$th tensor, considering the site index and the previous link index as the row indices, and the link index to the $b+1$th site as the column index.
    \item Calculate the entropy as $S_b = - \sum_j p_j\log(p_j)$, where $p_j$ is the square of the $j$th singular value.
\end{enumerate}

With this definition, for every eigenstate, there are $N-1$ different values of
entanglement entropy depending on how the system is partitioned. In
Fig.~\ref{fig:entanglement-entropy}, we demonstrate the entanglement entropy for
each state as a function of the bond number. In the body of the paper, we report
the average value of all of these numbers for every state.

\section{Dimeric Spectra Obtained from DMRG}

\begin{figure}
    \centering
    \subfigure[along $\vec{a}$]{\includegraphics{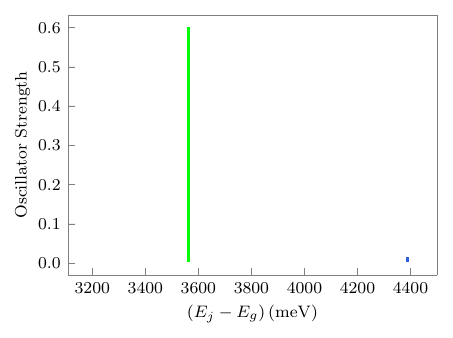}} 
    \subfigure[along herringbone direction]{\includegraphics{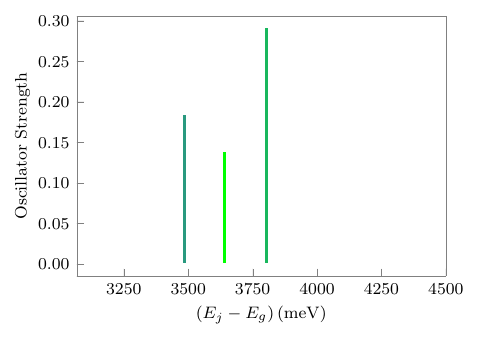}}
 \caption{Dimer spectra}\label{fig:dimer_spec_comp}
\end{figure}

We provide the dimer spectra for the dimers in this section. The excitation
energies and the oscillator strength are already given in the Table I and II in
the manuscript. Figure~\ref{fig:dimer_spec_comp} shows the comparison of the
spectra of the dimer along the two directions. The color of each stick also
signify the nature of the state. Because of mixing between LE and CT states in
case of herringbone dimer we see many peaks compared to only one peak in the
parallel dimer, where the states are not mixed. 
\bibliographystyle{unsrt}
\bibliography{ref}